\documentclass[manuscript]{aastex}

\slugcomment{Not to appear in Nonlearned J., 45.}

\shorttitle{Cusp-shaped structure of a Jet} \shortauthors{Zhang et
al.}

\begin{document}

\title{Cusp-shaped structure of a jet observed by \emph{IRIS} and \emph{SDO}}

\author{Yuzong Zhang and Jun Zhang}
\affil{Key Laboratory of Solar Activity, National Astronomical
Observatories, Chinese Academy of Sciences, Beijing 100012, China;
yuzong@nao.cas.cn; zjun@nao.cas.cn}

\begin{abstract}
 On 29 August 2014, the trigger and evolution of a cusp-shaped jet
 were
 captured in detail at 1330 {\AA} by the \emph{Interface Region Imaging Spectrograph}.
 At first, two neighboring mini-prominences arose in turn from low solar atmosphere and collided with a loop-like
 system over them. The collisions between the loop-like system and
  the mini-prominences lead to the blowout and then a cusp-shaped jet formed with a spire and an arch-base.
  In the spire, many brightening blobs originating from
  the junction between the spire and the arch-base, moved upward in a
   rotating manner and then in a straight line in the late phase of the jet.
    In the arch-base, dark and bright material simultaneously tracked in a fan-like structure and the
    majority of the material moved along the fan's threads.
     At the later phase of the jet's evolution, bidirectional flows emptied the
     arch-base, while down-flows emptied the spire, thus making the
     jet entirely vanish. The extremely detailed observations in this study shed new light on
      how magnetic reconnection alters the inner topological structure of a jet, and provide a beneficial complement for understanding
     current jet models.
\end{abstract}


\section{Introduction}
A solar jet is a highly dynamic, transient, and well collimated
eruption phenomenon, which is thought to contribute to coronal
heating and solar wind acceleration
\citep{1983ApJ...272..329B,2009ApJ...691...61P,2014Sci...346A.315T}.
With the soft X-ray telescope \citep{1991SoPh..136...37T} on board
the \emph{Yohkoh} satellite \citep{1991SoPh..136....1O},
\citet{1992PASJ...44L.173S} discovered X-ray jets, forcefully
motivating the jet study on observations and related theories.
Generally, the family members of jets, including spicule,
photospheric jet, H$\alpha$ surge, Ca II H jets, EUV jet, X-ray jet,
and macro spicule, can potentially be observed by all spectral lines
of present instruments with quite distinct morphology and structure
\citep{2010ApJ...714.1762P}, such as standard jets
\citep{1992PASJ...44L.173S,1995Natur.375...42Y}, blowout jets
\citep{2010ApJ...720..757M,2013ApJ...769..134M} and micro-sigmoid
jets \citep{2010ApJ...718..981R}. In principle, a jet is composed of
an arch-base and an elongated spire standing on the arch-base.

With the improvement of the observational capability on the solar
atmosphere, more new features on the jet geometry, dynamics, and
physical property are unceasingly revealed. For example, the
extensive rotating or spinning motions in the spire show an
untwisting mechanism
\citep{1996ApJ...464.1016C,2009SoPh..259...87N,2011ApJ...728..103L,2011ApJ...735L..43S,2012RAA....12..573C,2014ApJ...782...94L,
2015ApJ...801...83C,2015ApJ...798L..10L}. The upflowing motion along
the spire could even reach to a very high speed approximately 800 km
s$^{-1}$, which is near the local Alfv\'{e}n speed and this
suggested to possibly contribute to the high-speed solar wind in a
coronal hole \citep{2007Sci...318.1580C}.
 Recently, some recurrent helical jets were
reported
\citep{2007Sci...318.1580C,2010ApJ...714.1762P,2011A&A...531L..13I,2014Sci...346A.315T,2015ApJ...801...83C,2015ApJ...815...71C},
and well reproduced by numerical simulations on the oscillatory
magnetic reconnection between an emerging field and a pre-existing
active region field \citep{2010A&A...512L...2A}. Moreover, the
recurrent jets with untwisting motions and upflowing at near the
local Alfv\'{e}n speed were reproduced in the numerical experiment
by \citet{2015ApJ...798L..10L}, directly showing that the helical
jet is related to the torsional Alfv\'{e}n wave propagating in the
corona. In addition, a jet spire frequently appears whipping-like
motion, suggesting a relaxing process of the reconnected magnetic
field \citep{1995Natur.375...42Y,2007PASJ...59S.745S}. From the same
case in \citet{2007PASJ...59S.745S}, \citet{2014A&A...561A.134Z}
first observed the transverse oscillation of the spire. Furthermore,
the multi-temperature observations by different instruments prove
the jet temperature is not unique, comprising cool and hot
components
\citep{2000A&A...361..759Z,2007A&A...469..331J,2008ApJ...683L..83N}.
Additionally, with the higher spatial and temporal resolution
observations at EUV wavebands, some jets have been found to have
micro-CME (coronal mass ejection) like shape
\citep{2009SoPh..259...87N,2012ApJ...745..164S}, which hint a
scale-invariant eruptive phenomenon of the Sun
\citep{2016SSRv..tmp...31R}.

Based on the key mechanism of magnetic reconnection for impulsively
releasing energy, two main models on jet are very popular at
present. One is the standard model, which is proposed and
numerically simulated in two dimensions (2D), and 3D
\citep{1977ApJ...216..123H,1992PASJ...44L.173S,1995Natur.375...42Y,1996PASJ...48..353Y,1996PASJ...48..123S,2008ApJ...683L..83N,2009ApJ...691...61P}.
 In this model the closed magnetic arch emerges into the ambient
unipolar field and then reconnection occurs between the unipolar
field and one branch of the arch with an opposite polarity to the
unipolar field. The other is the blowout model
\citep{2010ApJ...720..757M,2013ApJ...769..134M} with two magnetic
reconnection processes. The first magnetic reconnection, which is
similar to that in the standard model, occurs between the emerging
magnetic arch and the ambient unipolar field. Then a filament with a
large amount of pent-up free energy by shear (or twist) dominates
the second magnetic reconnection
\citep{2011ApJ...735L..18L,2011ApJ...728..103L,2013ApJ...771...20M,2014ApJ...796...73H,2015ApJ...798L..10L}.
Recently, \citet{2015Natur.523..437S} suggested that all of the jets
in the coronal hole essentially come from mini-filament eruptions.
According to the latest review \citep{2016SSRv..tmp...31R}, the
standard jets are intensively collimated, usually form  an inverse-Y
shape with a narrow spire, and seldom rotate. However, the blow-out
jets strongly rotate, and posses a  relatively broad spire. As
suggested by \citet{2010ApJ...714.1762P}, a jet process is always
dominated by a synthetic role from magnetic geometrical
characteristics of the jet source, such as size, flux distribution,
null point position, and asymmetry. Many observations find that jets
could appear with similar morphology and outflow, but might possess
distinctive magnetic structures in the low coronal source region
 \citep{2009SoPh..259...87N,2012ApJ...746...19Z}.

In total, a large number of observations and theoretical models
basically describe and explain the morphology, dynamical and
physical properties of the jets. However, for the limitation of
observational capability, there is no appreciable resolution to
distinctly resolve the motions of the plasma blobs
\citep{2014A&A...567A..11Z} and related variation of magnetic
topological structure inside a jet. Fortunately, benefitting from
the extremely high spatial and temporal resolution in special
wavelength observations by the \emph{Interface Region Imaging
Spectrograph} \citep[{IRIS};][]{2014SoPh..289.2733D} and all
ultraviolet (UV) and extremely ultraviolet (EUV) wavelengths of the
Atmosphere Imaging Assembly \citep[{AIA};][]{2012SoPh..275...17L}
aboard the \emph{Solar Dynamics Observatory}
\citep[\emph{SDO};][]{2012SoPh..275....3P}, a cusp-shaped structure
jet was captured with unprecedented detailed observations. The
formation and evolution processes of the jet, including the motions
of the blobs and topological variation of its skeleton, are
described in Sec. 2. Section 3 contains discussion and conclusions.

\section{Observations}
\subsection{Data}

From 05:51 UT to 06:39 UT on 29 August 2014, \emph{IRIS} narrow-band
slit-jaw images (SJIs) (see Fig. 1) in the 1330 {\AA} channel
captured a cusp-shaped jet located at N08W90 (Solar-X: 935$\arcsec$
and Solar-Y: 185$\arcsec$) in detail. The field of view (FOV) of
each SJI 1330 was 119$\arcsec$$\times$119$\arcsec$, but total SJIs
1330 covering a larger Solar-Y range of 133$\arcsec$ by scanning the
solar surface from north to south with an eight-step round and a
2$\arcsec$ step length. The jet was located exactly at the
northernmost region of the larger FOV; therefore, a data gap in the
north fraction of the jet with a width range of
2$\arcsec$$\--$14$\arcsec$ was unavoidable while the FOV of the SJIs
1330 left the northernmost region. However, the main jet process was
still successfully recorded. As a complement, all the UV and EUV
data observed by \emph{SDO}/AIA were adopted to display the
different temperature components of the jet. The time cadence and
the spatial resolution of the SJIs are 9.6 s and 0.166$\arcsec$ per
pixel, respectively, while those of the images observed by the
\emph{SDO}/AIA are 12 s and 0.6$\arcsec$ per pixel, respectively.
Furthermore, to ease reading, all of the images were rotated
90\symbol{'027} counterclockwise, i.e., the left and right represent
the original north and south, respectively. Finally, a smaller FOV,
with a Solar-X size of 60$\arcsec$ (from 991$\arcsec$ to
931$\arcsec$) and a Solar-Y size of 38$\arcsec$ (from 144$\arcsec$
to 182$\arcsec$) was cut to merely focus the jet.

\subsection{Formation and evolution of the cusp-shaped jet from \emph{IRIS} observations}
At 05:51:04~UT on 29 August 2014 a closed-loop-like structure
(hereafter, ``loop''; and marked ``L1'' in Fig. 1) appeared in the
\emph{IRIS} SJI 1330 and then slowly ascended with a speed of
approximately 15 km s$^{-1}$. Five minutes later, a mini-prominence
(see ``P1'' in Fig. 1) with an uneven brightness structure appeared
under the loop and ascended with a higher speed of approximately 65
km s$^{-1}$. Affected by the newly appearing mini-prominence, the
loop accelerated to a much higher speed of approximately 50 km
s$^{-1}$. One minute after its appearance, P1 expanded violently and
collided with L1 (see Figs. 1C and D). In the following 2 min, L1
covered and prevented P1 from freely expanding and erupting,
resulting in both of them moving up together at a speed of about 60
km s$^{-1}$. Simultaneously, many small cusp-shaped structures
appeared at the common apex of L1 and P1 (e.g., the ``Cusp 1'' in
panel E of Fig. 1). These small cusp-shaped structures contributed a
tunnel for plasma blobs flowing out from the erupted
mini-prominence. Then these small cusp-shaped structures developed
and merged into one cusp-shaped structure with an outstanding spire
at the apex of its arch-base. Later, another pair of mini-prominence
(P2) and loop-like system (L2) also collided and formed another
cusp-shaped structure (see Figs. 1G-I). Finally, the above two
cusp-shaped structures combined into a complete jet structure with a
notable cusp-shaped structure composed of a spire and an arch-base.
The width of the arch-base is mostly the same as that of L1 when L1
collided with P1. Therefore, the jet in this event consists of two
sets of similarly cusp-shaped structures with some distinction in
scale. Totally, it merely took 9 minutes for the formation of the
jet since the first loop appeared.

Subsequently, a huge number of bright plasma blobs (e.g., shown by
the diamonds in Fig. 2A) appeared and traced in detail the
topological evolution of the jet. These plasma blobs first moved up
from the bottom of the left leg of the arch-base. As they arrived at
the junction of the spire and the arch-base, they started to
bifurcate and formed a bidirectional flow. The solid blue arrows in
Fig. 2A show one branch of the bidirectional flow moving up along
the spire, while the dashed blue arrow denotes the other branch
moving down along the right leg of the arch-base. For the plasma
continuously discharging from the left leg of the arch-base, the
original thick left leg degraded to a clear skeleton, which is
composed of three thin legs (see Fig. 2B). Then bright plasma blobs
(see Figs. 2D-E) were ceaselessly flowing up/down along these thin
legs, subsequently these legs became thinner. Some details are
illustrated in Figs 3A-D, e.g., a couple of brightening blobs
appeared at about 06:03:03 UT, and then immediately separated. One
blob (marked by solid circles) moved downwards with the speed of
approximately 64 km s$^{-1}$, meanwhile the other one (marked by
asterisks) was elongated and ascended also along the thin leg.
Additionally, the size of an identified smallest blob in this jet
reaches to about 0.8$\arcsec$ (see the blob denoted by the long
green arrow passing through Figs. 2D and E). These thinner legs
became extreme thin and then completely broke at approximately
06:06:53 UT (see Fig. 2F) resulting in the arch-base disappearance,
hence only the spire of the jet remained.

As shown in Figs. 2G-H and more details in Figs. 3E-H, continually
brightening processes led to the appearance of up/down-flows near
the elbow of the remaining bending spire. The up-flows were
apparently accelerated, as well as the down-flows pushed powerfully
down the spire elbow. At about 06:09:26~UT, a violent ``explosion''
occurred at the left edge of the elbow, resulting in some plasma
blobs promptly moving upwards with a speed of approximately 163 km
s$^{-1}$, while others moving down. Later, the productivity of the
brightening processes gradually slowed down. After 06:20:46~UT,
while the brightening processes finally ceased in the jet, an amount
of plasma began to directly flow back to the solar surface along the
spire. Then the spire grew fainter, and completely vanished at about
06:39:26~UT.

\subsection{Complementary \emph{SDO}/AIA observations of the jet}

To compensate the missing data of the jet in the \emph{IRIS} SJIs
1330, we adopted a total of nine wavelength images in the UV and EUV
captured by the \emph{SDO}/AIA as complementary observations. We
found that the basic configuration and evolution process of the jet
observed by both solar satellites are quite similar. However, for
distinct temperature responses at different wavelengths, there are
outstanding observational distinctions in detail. Regarding the
observations of the first pair of the loop and the mini-prominence
by AIA, only a fraction of L1 and most of P1 could be captured at
1600 {\AA}. No loop could be seen, but dark P1 appeared at 193
{\AA}. A part of L1 and P1 could be observed at 335 {\AA}. Figures
4E-I display both L1 and P1 as a dark region at 304 {\AA}. During
the evolution stage of the jet, much dark matter was observed at 304
{\AA} to escape from P1 and almost float around the whole jet. In
addition, the bright and dark matter traced the fan-like skeleton of
the arch-base.

At 171 {\AA} P1 displayed as a dark feature before colliding with
L1. After the collision, several bright points
 appeared in dark P1. Gradually, the arch-base
structure became more and more distinctive amidst the ambient dark
mini-prominence material. With most of the dark material moving
around the spire, the remainder was finally heated as the bright
ingredients of the arch-base. During both the formation and
evolution processes of the cusp-shaped jet, many mini-bright-points
appeared in the left branch of the arch-base. Four bright points and
their motion processes are clearly presented in Fig. 5. After the
collision between L1 and P1, the first bright point (cf. Figs. 5A
and B) appeared, moved up and was elongated into a bright
thread-like structure in the formation process of the spire. The
first one and the following two bright points (see Figs. 5A-F),
spanning both the formation and the evolution processes of the jet,
are found outstanding rotation along the spire. However, after the
arch-base disappeared for the sudden collapse of the thinner legs
mentioned in Fig. 2F, the rotation of the spire slowed down.
Therefore, the fourth bright point shown in Figs. 5G and H, just
moved along a thread-like structure in the spire without any obvious
rotating behavior. Additionally, after the last bundle of light
material out-poured up from the elbow of the remaining spire at
06:23:29~UT (cf. Fig. 2I), all of material began to fall down along
the spire without any apparent rotation.

\section{Discussion and conclusions}
In this study, from the observations by \emph{IRIS} SJIs 1330 and
\emph{SDO}/AIA in all the UV and EUV wavelengths with unprecedented
spatial and temporal resolution, we are offered a chance to learn
about the formation of a cusp-shaped jet and the dynamics of the
material in the jet. In the previous literatures, it is seldom
reported that a cusp-shaped jet with a broad spire and an arch-base
forms by the collisions between loop-like system and
mini-prominences. A lot of brightening blobs moved upward along the
spire in a rotating manner, and some material moved downward along
the arch-base. Finally, the bidirectional flows emptied the spire
and the arch-base, thus making the jet vanish.

\citet{2010ApJ...720..757M,2013ApJ...769..134M} proposed dichotomy
of solar coronal jets, i.e. standard jets and blowout jets. Their
observational samples show a cool component in all blowout jets and
in a minority of standard jets, outstanding lateral expansion in the
former but none in the latter, and obvious axial twist in both
classes of jets. Recently,
\citet{2015Natur.523..437S,2016ApJ...821..100S} suggested that both
standard and blowout jets are related to the filament eruptions but
in different explosion intension. In this event, the collision
between the loop-like system and the mini-prominence leads to the
blowout and then the jet forms. The cool component, lateral
expansion and obvious axial rotation of the spire of the jet are all
observed. These properties are consistent with the scenarios
described by them. However, besides of the mini-prominence
eruptions, the loop-like system also takes part in the jet
formation. The loop-like system keeps the mini-prominence from free
eruption and forms a fine arch-base together with the
mini-prominence. Therefore, this event basically fits into the
scenarios described by
\citet{2010ApJ...720..757M,2013ApJ...769..134M} and
\citet{2015Natur.523..437S,2016ApJ...821..100S} with new features,
 which are mainly benefited from the extremely high observational
resolution by IRIS. Totally, the jet appears all of the typical
observational features of the blowout jet (e.g., an outstanding
rotation, a relatively broad spire and a mini-filament eruption)
reviewed by the latest literature
 \citep{2016SSRv..tmp...31R} and completely fits with the class of
blowout jet.

The closed loop-like system in this case is easily reminiscent of
another two jets reported by \citet{2009ApJ...707L..37L} and
\citet{2014ApJ...782...94L}. In particular , a distinct loop is also
observed in \citet{2009ApJ...707L..37L}. In the jets studied by
them, the spires of the jets show evidently helical motions, which
denote the untwisting magnetic field lines and are usually supposed
to be a characteristic of a blowout jet. Additionally, both of their
spires appear interesting transverse motions or wriggling
perpendicular to the jet axis. However, the jet described in this
paper shows that the spire gradually formed by a series of small
cusp structures which merge and grow without outstanding transverse
motions. Nevertheless, thanks to the high spatial resolution and
suitable observational angle , the detailed formation and evolution
processes of the arch base and the plasma blob motions are clearly
captured, which might shed new light into the physical mechanism in
the interior of a jet.

Among these small-scale brightening plasma blobs, their smallest
size even reaches to 0.8$\arcsec$, which is consistent with that of
the observations in Ca II H by the Solar Optical Telescope
\citep[SOT, ][]{2008SoPh..249..167T,2012ApJ...759...33S} aboard
\emph{Hinode} \citep{2007SoPh..243....3K}, but much smaller than
that in observations by \emph{SDO}/AIA \citep{2014A&A...567A..11Z}.
These brightening blobs trace the bidirectional flows forming at the
junction of the arch-base and spire, the variation of the rotation
velocity of the spire, and the detailed evolution in the fan-like
branches of the arch-base. Once all of the brightening blobs
disappear in the late phase of the jet, an enormous amount of
material immediately returns back to the solar surface, resulting in
the final disappearance of the jet.

Significantly, we observed many brightening plasma blobs (see Figs.
2D-E) and bidirectional flow \citep{1997Natur.386..811I} traced by
these blobs (see Figs. 3A-D) in the interior of the jet.  These
observational evidence suggests that there exist plenty of magnetic
reconnection processes. In theory, according to the VAL-C model
\citep{1981ApJS...45..635V,1993ApJ...406..319F} for the upper
chromosphere, the magnetic Reynolds number is quite outstanding on
the order of 10$^{6}$, which easily leads to the magnetic island
appearance during the magnetic reconnection process
\citep{2009PhPl...16k2102B}. Moreover , in the recent numerical
experiments on the jet process
\citep{2013ApJ...777...16Y,2015ApJ...799...79N}, it is found that
comparing to their plasma environment, the magnetic islands always
possess relatively higher temperature and density, so appear more
bright. Therefore, these brightening blobs, which could be used to
trace the bidirectional flows, are extremely possibly magnetic
islands occurring in the process of magnetic reconnection for the
tearing-mode instability
\citep{1963PhFl....6..459F,2000A&A...360..715K,2004ApJ...605L..77A,2005ApJ...622.1251L,2012ApJ...758...20N,2013A&A...557A.115K,2014A&A...567A..11Z}.
This hints that there are complex and ubiquitous magnetic
reconnections of different scales in the entire jet process to heat
and transfer the material, as well as alter the magnetic topological
structure.

\acknowledgments

\emph{IRIS} is a NASA Small Explorer mission developed and operated
by LMSAL with mission operations executed at NASA ARC and major
contributions to downlink communications funded by the NSC (Norway).
\emph{SDO} data and images are courtesy of NASA/\emph{SDO} and the
AIA and HMI science teams. The work is supported by the National
Natural Science Foundation of China (11533008, 11303049, 11673034,
 and 11673035), the Strategic Priority Research Program-The Emergence
of Cosmological Structures of Chinese Academy of Sciences (No.
XDB09000000).

\clearpage



\begin{figure}
\epsscale{0.75} \plotone{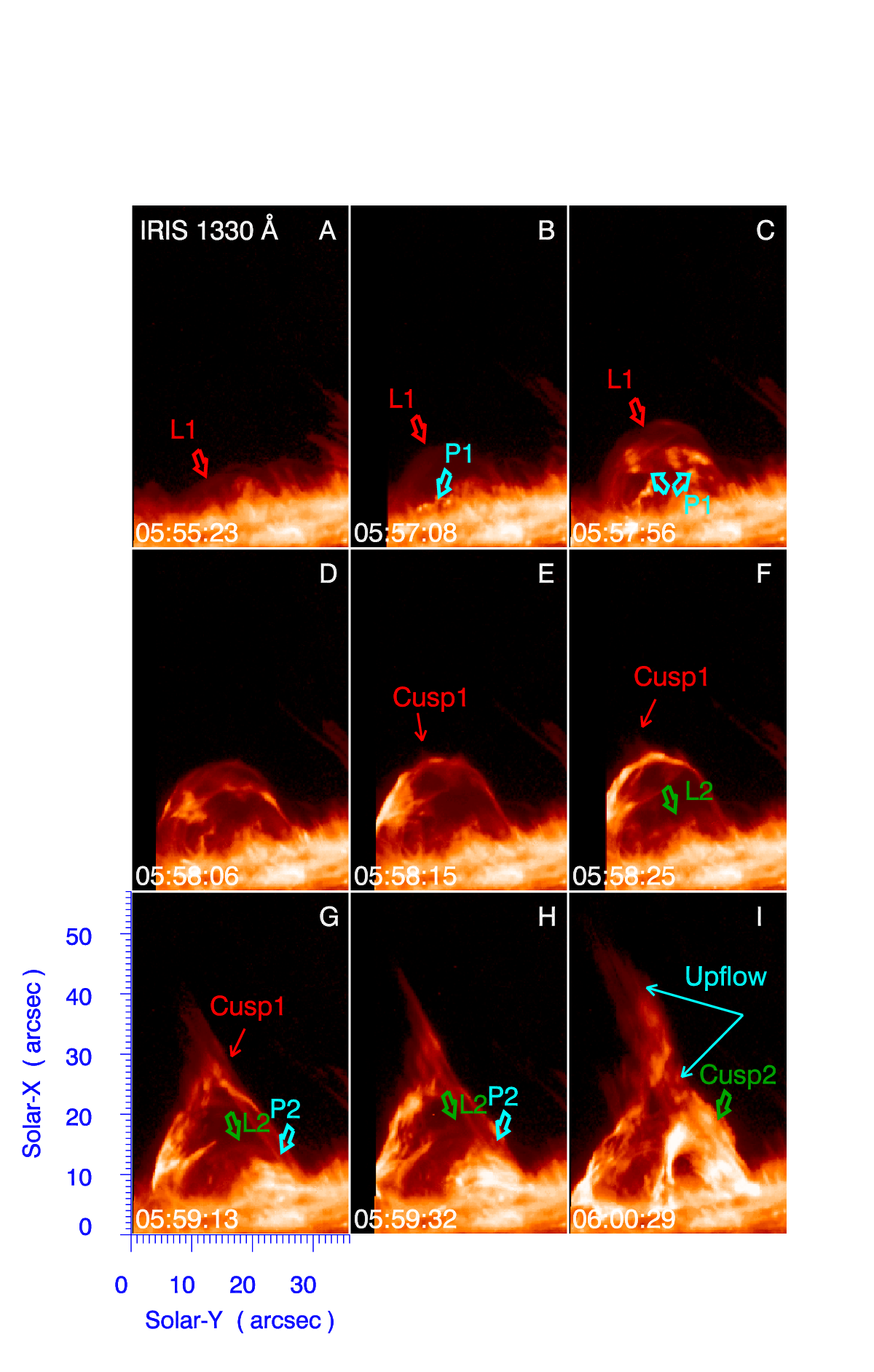} \caption{Formation process of
the cusp-shaped jet observed in the \emph{IRIS} SJIs 1330. ``L1''
and ``L2'' denote the first and second closed-loop-like objects;
``P1'' and ``P2'' show the first and second mini-prominences. The
jet is composed of two cusp-shaped structures ``Cusp 1'' and ``Cusp
2'' overlaid on the line of sight. ``Upflow'' shows the plasma
flowing upwards along the spire of the jet. \label{fig1}}
\end{figure}

\clearpage

\begin{figure}
\epsscale{0.7} \plotone{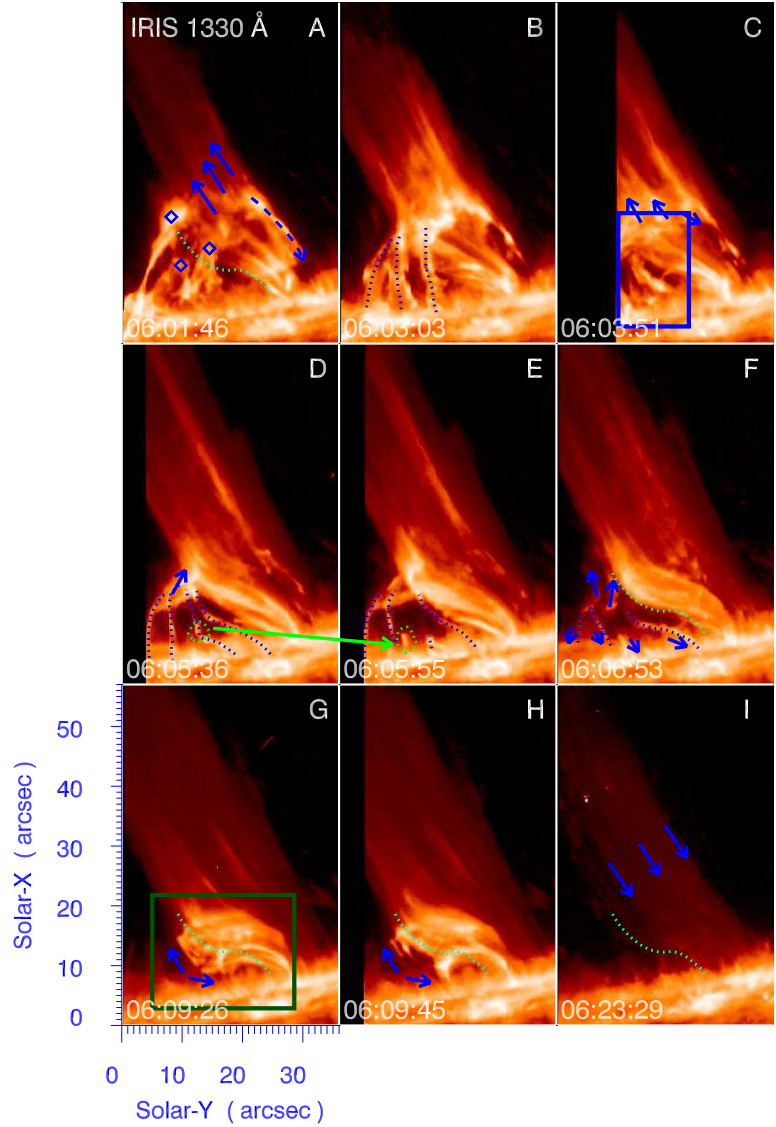} \caption{Evolution process of
 the cusp-shaped jet. Diamonds mark the jet plasma blobs. Blue arrows denote the
motion directions of the jet plasma at different times. Light green
dotted line shows the lower edge of the spire at 06:06:53 UT. Blue
dotted lines mark the fine branches of the left leg of the
arch-base. Long green arrow links the corresponding plasma in the
green dotted circles in panels D and E, respectively. Motions of the
plasma in blue and green boxes are described in detail in Fig. 3.
\label{fig2}}
\end{figure}

\clearpage

\begin{figure}
\epsscale{0.7} \plotone{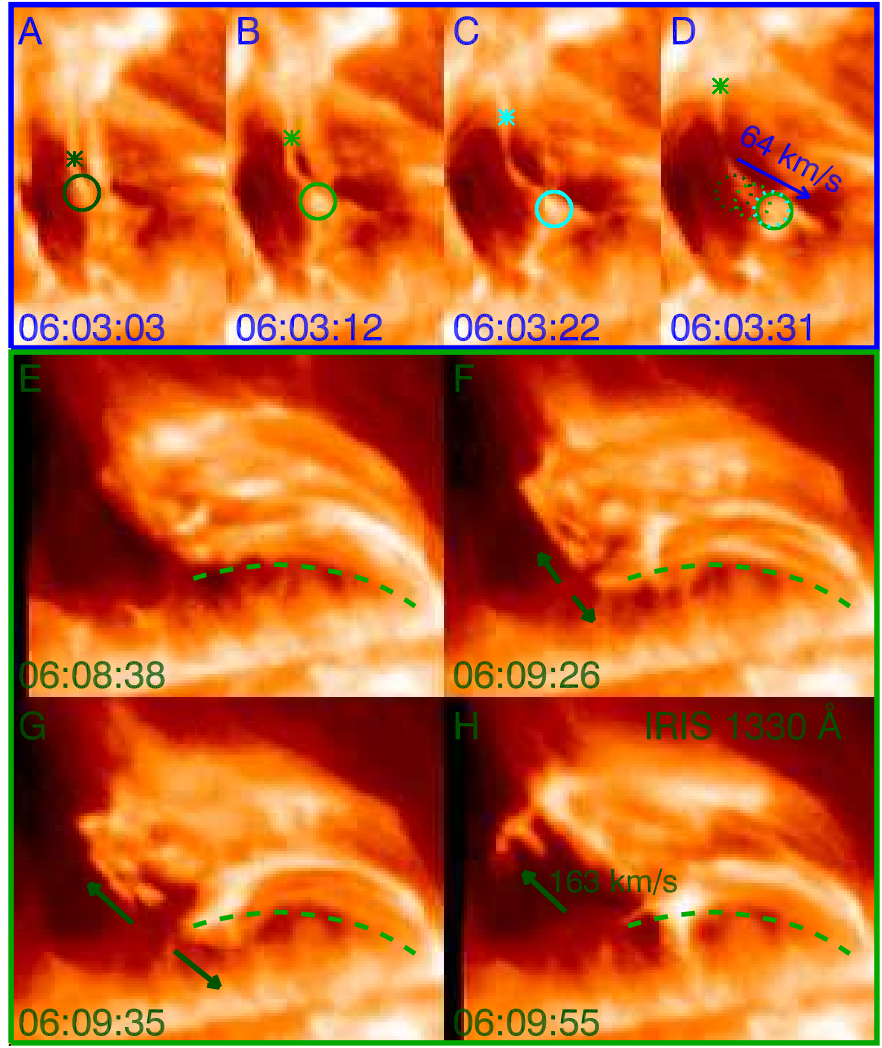} \caption{Two examples of the
plasma motions during the jet evolution process captured by the
\emph{IRIS} SJIs 1330. Successive images, encircled in blue and
green boxes, correspond to those images in blue and green boxes in
Fig. 2, respectively. Circles and asterisks in the blue box denote
two plasma blobs moving in the opposite directions. Arrows trace the
motions of the bright points. Green dashed curve line marks the
lower edge of the spire at 06:08:38 UT on 29 August 2014.
\label{fig3}}
\end{figure}

\clearpage
\begin{figure}
\epsscale{0.8} \plotone{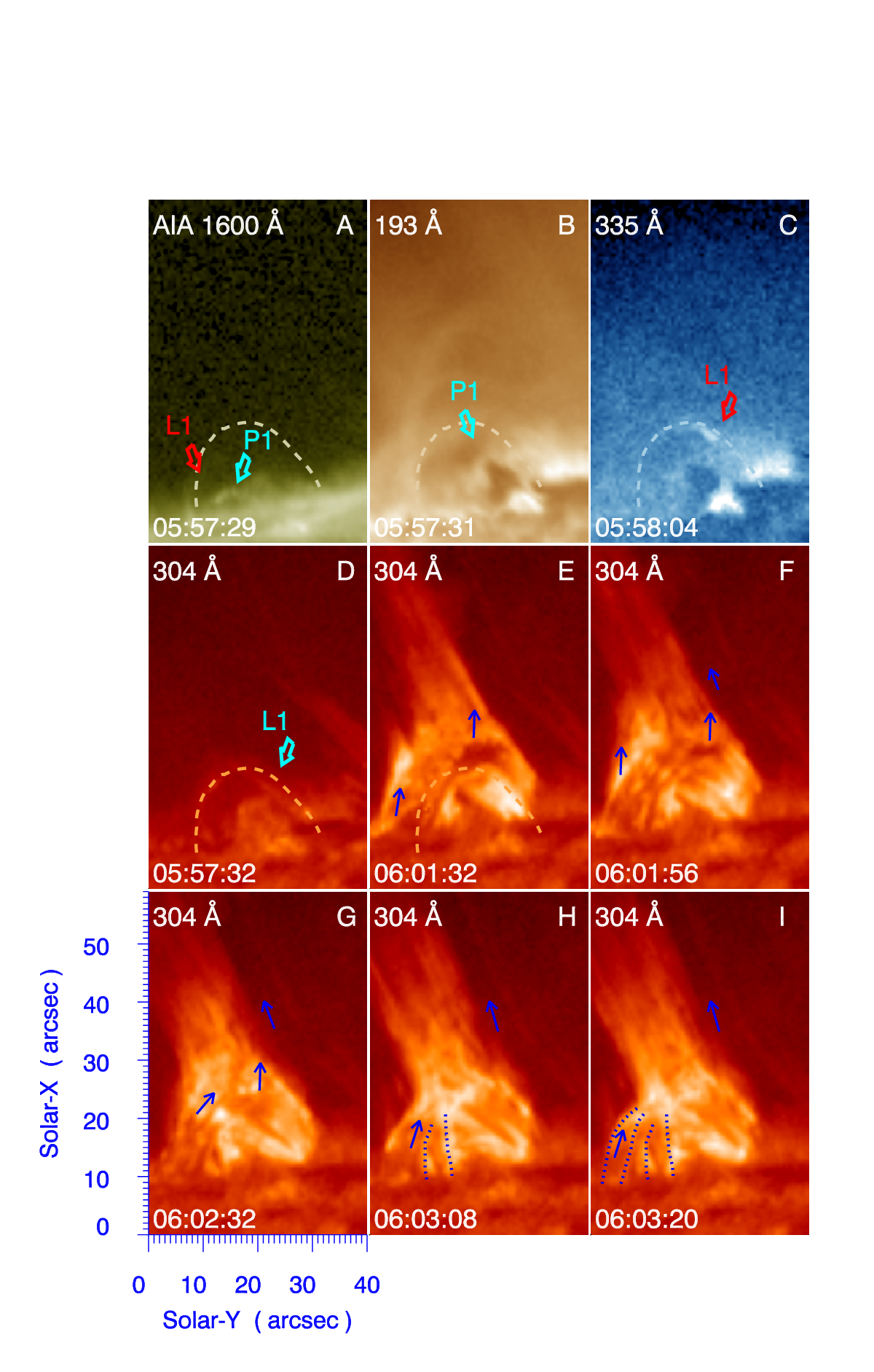} \caption{Jet observed in four
wavelengths by \emph{SDO}/AIA. Dashed line denotes the first
closed-loop-like structure in the \emph{IRIS} SJIs 1330 at
approximately 05:57:56 UT. ``L1'' and ``P1'' are the same as those
in Fig. 1. Blue dotted lines mark the fine branches of the left leg
of the arch-base. Small blue arrows mark the dark plasma that
escaped from the mini-prominence.\label{fig4}}
\end{figure}

\clearpage

\begin{figure}
\epsscale{1.0} \plotone{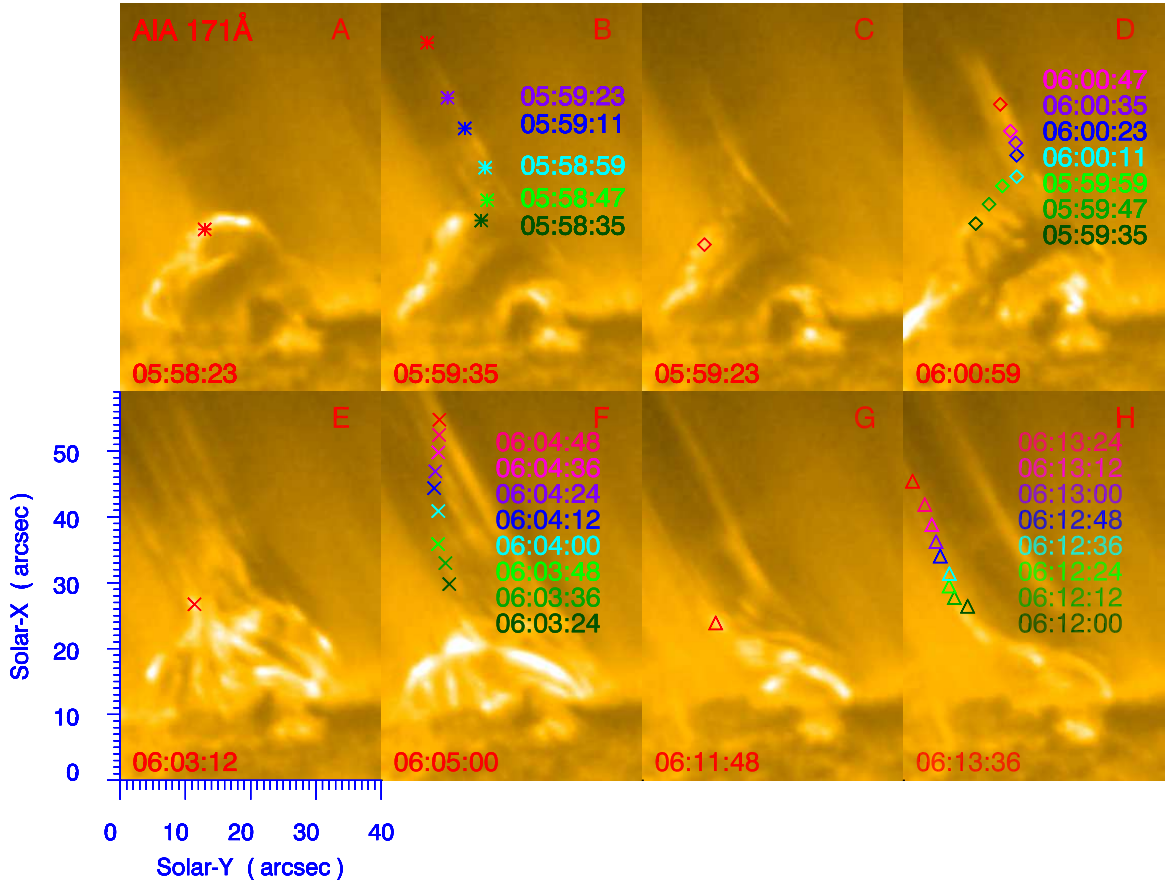} \caption{Tracking the motions
of four brightening points originating from the apex of the
arch-base in the successive \emph{SDO}/AIA 171 {\AA} observations.
Traces of each brightening point are marked by the same symbols.
Time of each position is written in the same color of the
corresponding symbol.\label{fig5}}
\end{figure}

\clearpage

\end{document}